\documentclass[a4paper, prl, twocolumn, showpacs]{revtex4}

\usepackage{amssymb}
\usepackage{amsmath}
\usepackage{color}
\usepackage{graphics, graphicx}
\usepackage{bbold}
\usepackage{psfrag}
\usepackage{mathcomp}

\begin{document}

\author{I. Titvinidze}
\author{M. Snoek}
\author{W. Hofstetter}
\affiliation{Institut f\"ur Theoretische Physik, Johann Wolfgang Goethe-Universit\"at, 60438 Frankfurt am Main, Germany}

\title{Supersolid Bose-Fermi Mixtures in Optical Lattices}

\begin{abstract}
We study a mixture of strongly interacting bosons and spinless fermions with on-site repulsion in a three-dimensional optical lattice. For this purpose we develop and apply a generalized DMFT scheme, which is exact in infinite dimensions and reliably describes the full range from weak to strong coupling. We restrict ourselves to half filling. For weak Bose-Fermi repulsion a supersolid forms, in which bosonic superfluidity coexists with charge-density wave order. For stronger interspecies repulsion the bosons become localized while the charge density wave order persists. The system is unstable against phase separation for weak repulsion among the bosons.
\end{abstract}

\maketitle
Ultracold atomic gases confined in optical lattices provide a new laboratory  
for investigating quantum many-body problems with high precision and tunability \cite{Jaksch,Greiner}. 
In this way new light can be shed on notoriously difficult problems in condensed matter physics \cite{Hofstetter02}.  
One of the intriguing aspects of cold atoms is that the atomic quantum statistics can be controlled. In particular, cold atomic gases offer the possibility to realize mixtures of fermions and bosons \cite{Truscott, Hadzibabic2,Roati,Silber,Zaccanti}, recently also in an optical lattice \cite{Sengstock,Guenter}. 
It is thus possible to create physical systems without analog in conventional solid state physics. 

An exotic quantum phase that has intrigued researchers for a while   
is the \emph{supersolid} with superfluid order, i.e. broken $U(1)$ symmetry, and coexisting 
particle density wave order. It is still an open question whether a supersolid has been realized 
in recent experiments on $^4$He \cite{supersolid}. 
While in single-component quantum gases supersolids can only be stabilized by including nearest neighbor repulsion between the particles
\cite{supersolid_single_component},  they can be 
conveniently realized in Bose-Fermi mixtures with on-site repulsion as we show in this Letter. 
Earlier theoretical studies already suggested that Bose-Fermi mixtures can be unstable against 
charge density wave (CDW) and supersolid order or phase separation (PS). 
However, so far all theoretical approaches either dealt with one-dimensional systems \cite{Pollet, Mathey1,Imambekov, Mathey2, Roth, Cazalilla,Albus}, or relied on weak-coupling approximations \cite{Buechler,Lewenstein, Klironomos}. 

\begin{figure}
\includegraphics[scale=0.28]{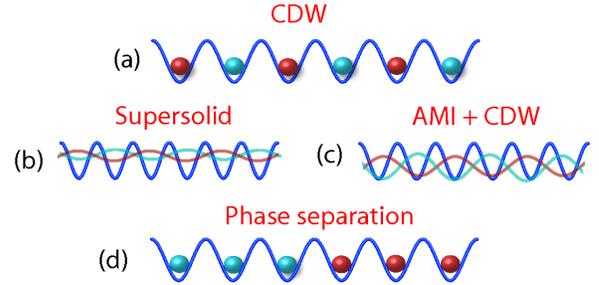}  
\caption{(Color online) Schematic structure of different \mbox{$T=0$} phases of a spinless Bose-Fermi mixture in an optical lattice. 
The red (dark) particles correspond to the fermions, while cyan (light) particles denote the bosons. In both the supersolid and the AMI phase the bosons and fermions have an alternating density pattern as depicted in (a). In the supersolid (b) the density oscillations are small and the bosons are superfluid. In the AMI + CDW phase (c) the density oscillations are large and the bosons are localized. The schematic 
structure of phase separation is depicted in (d).
}

\label{phases}
\end{figure}

Here we introduce and apply a generalized dynamical mean-field theory (GDMFT)  
that treats this problem in a fully non-perturbative way. 
In this method the fermions are described by dynamical mean field theory (DMFT) \cite{DMFT}, into which the bosons are incorporated by means of the static Gutzwiller decoupling approximation \cite{Gutzwiller} of the hopping. 
This approach therefore reproduces the strong coupling behavior of both the fermions and the bosons. In particular it is able to describe the formation of a bosonic Mott insulator state for strong repulsion between the bosons at integer filling. 
Here we restrict ourselves to half filling for both the bosons and fermions ($ \langle n_b \rangle  = \langle  n_f  \rangle = \frac{1}{2}$). For weak interspecies repulsion we predict the formation of a {\em supersolid phase}, in which the bosons form a superfluid with spatially modulated density and the fermions form a CDW. For stronger repulsion between the bosons and the fermions there is a first order phase transition to an {\em alternating  Mott insulator} (AMI) plus CDW, in which the bosons become localized at every second lattice site while the fermionic charge density wave persists. For weak bosonic repulsion we find an instability towards {\em phase separation} (PS). We depict all these different phases schematically in Fig. \ref{phases}.

A mixture of fermions and bosons in an optical lattice can be described by the single-band Fermi-Bose Hubbard model:
\begin{eqnarray}
\label{Hamiltonian}
{\cal H}&=&-\sum_{\langle i,j \rangle}\left\{ t_{f}c_{i}^{\dagger}c_{j}+t_{b}b_{i}^{\dagger}b_{j} \right\}
-\sum_{i}\left\{\mu_{f}n_{i}^{f}+\mu_{b}n_{i}^{b}\right\}
\nonumber\\
&&+\sum_{i}\left\{\frac{U_{b}}{2}n^{b}_{i}(n^{b}_{i}-1)+U_{fb}n^{b}_{i}n^{f}_{i}\right\},
\end{eqnarray}
where $c_{i}^{\dagger}$ ($b_{i}^{\dagger}$) is the fermionic (bosonic) creation operator at site $i$, while  
$n^{f}_{i}= c_{i\sigma}^{\dagger}c_{i\sigma}$  ($n^{b}_{i}= b_{i}^{\dagger}b_{i}$) denotes the number operator 
and $\mu_{f(b)}$ the chemical potential for fermions (bosons).  $U_{b}$ and 
$U_{fb}$ are the on-site boson-boson and fermion-boson interactions respectively.  $\langle i,j \rangle$ 
denotes summation over nearest neighbors, and $t_{f(b)}$ is the tunneling amplitude for fermions (bosons). 

Following the very succesful DMFT \cite{DMFT} and Gutzwiller \cite{Gutzwiller} schemes, which are exact in infinite dimensions, we consider the infinite-dimensional limit ($d \rightarrow \infty$)  of this model first, which is expected to be a good approximation to three spatial dimensions.  
In order to retain a finite kinetic energy  the hopping parameters are rescaled as $t_f/\sqrt{d}$ and $t_b/d$. We then follow the standard ``cavity derivation'', i.e., we consider a single impurity site and integrate out all other sites \cite{DMFT}. In the limit of infinite dimensions the only terms that survive in the effective action for the impurity site are the local terms, plus a bosonic source field and a fermionic (Weiss-) mean-field. The bosonic part corresponds to the Gutzwiller approximation, whereas the fermionic part corresponds to DMFT. Therefore, the GDMFT employed in our calculation consists of the DMFT algorithm 
for the fermions \cite{DMFT}, combined with Gutzwiller mean-field theory for the bosons. 
Subsequently the action for the impurity site is mapped onto a generalized single impurity Anderson model (GSIAM). 
As usual, the impurity site is coupled 
to a noninteracting fermionic bath, which provides a self-consistent dynamical (Weiss-) mean field \cite{DMFT}. 
In addition, the GSIAM now also contains a bosonic degree of freedom, which is self-consistently coupled to the superfluid order parameter, according to Gutzwiller mean-field theory \cite{Gutzwiller}. 
In summary, the GSIAM is described by the following Hamiltonian, which allows for a two-sublattice structure:
\begin{eqnarray}
&&\hspace{-6mm}\mathcal{H}_{\rm GSIAM}=\sum_{\sigma=  \pm 1} \bigl[\mathcal{H}_{b}^{\sigma} +  \mathcal{H}_{fb}^{\sigma} + \mathcal{H}_{f}^{\sigma}\bigr]   \label{GSIAM}  \\
&&\hspace{-6mm}\mathcal{H}_{\rm b}^{\sigma} = - z t_b (\varphi_{\bar \sigma} b^\dagger_{\sigma} + 
\varphi^{*}_{\bar\sigma} b_{\sigma}) + \frac{U_{b}}{2} n^{b}_{\sigma} (n^{b}_{\sigma} -1) - \mu_b  n^b_{\sigma} 
\nonumber  \\
&&\hspace{-6mm}\mathcal{H}_{\rm fb}^{\sigma} = U_{fb} n_{\sigma}^f  n_{\sigma}^{b} \nonumber  \\
&&\hspace{-6mm}\mathcal{H}_{f}^{\sigma} = - \mu_f n^{f}_{\sigma} +
\sum_{\bf k}\Bigl\{  \varepsilon_{\bf k} a^{\dagger}_{{\bf k}\sigma}a_{{\bf k}\sigma} 
+V_{{\bf k}\sigma}\left(c^{\dagger}_{\sigma}a_{{\bf k} \sigma}+h.c.\right)\Bigr\}   \nonumber
\end{eqnarray}
Here $\sigma$ is the sublattice index ($\bar \sigma =  - \sigma$),  $z$ is the lattice coordination number, 
$\varphi_{\sigma}= \langle b_{\sigma} \rangle$ is the superfluid order parameter, 
and $V_{{\bf k}\sigma}$ are the fermionic hybridization matrix elements.  The hybridization function is defined as $\Delta_{\sigma}(\omega) = \pi\sum_{\bf k}|V_{{\bf k}\sigma}|^{2}\delta(\omega-\varepsilon_{\bf k})$. For convenience we perform our calculations on the Bethe lattice, which has a semielliptic non-interacting density of states $\rho(\varepsilon)=2 \sqrt{D^{2}-\varepsilon^{2}}/\pi D^2$. Here $D =2 \sqrt{z} t_f $ is the non-interacting fermionic half-bandwidth. In the following we take $D$ as the unit of energy.
The fermionic DMFT self-consistency relation on the Bethe lattice has the form
$\Delta_{\sigma}(\omega)= \frac{\pi}{4} A_{\bar \sigma} (\omega)$ \cite{DMFT}, 
where $A_\sigma (\omega)$ is the local fermionic interacting (impurity) spectral function.
To calculate $A_\sigma (\omega)$ and $\varphi_\sigma = \langle b_{\sigma} \rangle$ 
from the GSIAM (\ref{GSIAM}) 
we use the nonperturbative numerical renormalization group (NRG) technique \cite{NRG, params}.
The resulting hybridization function as obtained via the  spectral function $A_\sigma (\omega)$
and  the bosonic order parameter $\varphi_{\sigma}$ determine the new coefficients of the GSIAM. 
This procedure is iterated until convergence is reached. 
The GDMFT approach incorporates the local correlations between bosons and fermions in a fully non-perturbative fashion 
and thus reliably describes the full range from weak to strong coupling. 
In our calculations we use a cut-off for the number of bosons on the impurity site, which can 
be kept low due to the repulsive interactions, which suppress multiple occupancy of the bosons. 
All of the results presented here are obtained at $T=0$.

The self-consistent GDMFT procedure as described above can yield multiple stable solutions. 
To find the ground state of the system, we need to compare the energy of these solutions, which is given by 
\begin{equation}
\label{GSE}
\hspace{-0.15cm}\frac{E}{N}=\frac{1}{N}{\cal H}_{K}+ \frac{1}{2}\hspace{-0.15cm} \sum_{\sigma = \pm 1}\hspace{-0.15cm} \left( U_{fb} \langle n^{f}_{\sigma}n^{b}_{\sigma}\rangle + \frac{U_{b}}{2} \langle n^{b}_{\sigma}(n^{b}_{\sigma}-1)\rangle\right)
\end{equation}
where ${\cal H}_{K}/N=\tfrac{1}{2}\sum_{\sigma}\langle \mathcal{H}_{f}^{\sigma}\rangle -  z t_b \varphi_{-1}\varphi_{1}$ and 
the indices $\pm 1$ correspond to the two different sublattices.  
To calculate the fermionic part of the kinetic energy 
we use the same approach as for an
antiferromagnetic state, which also has a two-sublattice structure \cite{DMFT,Zitzler}:
\begin{eqnarray}
\label{Kinetic}
\frac{1}{N}{\cal H}_{K}=- z t_b \varphi_{-1}\varphi_{1} + \int_{-\infty}^{\infty}d\varepsilon \; \varepsilon \rho(\varepsilon)
\int_{-\infty}^{0}d\omega  A(\varepsilon,\omega),  
\end{eqnarray}
where $\rho(\varepsilon)$ is the fermionic non-interacting density of states and 
$A(\varepsilon,\omega)= - \frac{1}{\pi}\Im m\frac{1}{\sqrt{\zeta_{-1}\zeta_{1}}-\varepsilon}$
is the spectral function, with $\zeta_{\sigma}=\omega+\mu_{f}-\Sigma_{\sigma}(\omega)$ ($\sigma = \pm 1$). We calculate the self-energy following \cite{Bulla}:
\mbox{$ \Sigma(\omega)=U_{fb} F(\omega) / G(\omega) $}
where $G(\omega)$ is the fermionic Green's function and $F(\omega)=\langle f b^{\dagger}b f^{\dagger}\rangle_{\omega} $.


\begin{figure}
\includegraphics[scale=0.28, angle=-90]{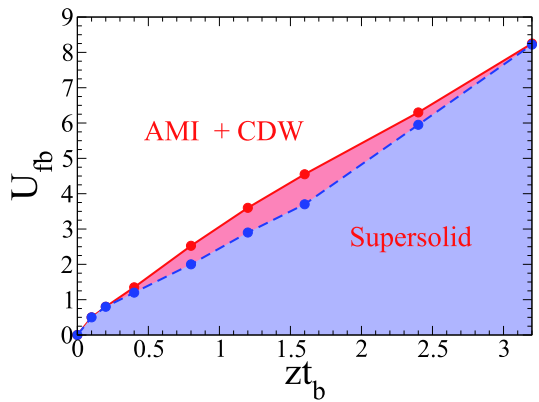}
\caption{(Color online) Phase diagram of the Fermi-Bose Hubbard model with spinless fermions and hard-core bosons 
at half filling. We identify the supersolid phase (below red solid line), the alternating Mott insulator (AMI) phase with charge 
density wave (CDW) (above red line), and the coexistence region (between the  red and blue (dashed) line). Energies are expressed in units of the non-interacting fermionic half-bandwidth $D$.}
\label{PD_hard_core}
\end{figure}

We now first apply our GDMFT procedure to the limit $U_{b}=\infty$, i.e. hard-core bosons. We fix the number of bosons and fermions at half filling ($ \langle n_b \rangle  = \langle  n_f  \rangle = \frac{1}{2}$), which makes the system particle-hole symmetric. 
Without loss of generality, calculations are performed for repulsive Fermi-Bose interactions: $U_{fb}>0$. The case of attractive interactions will be inferred later on with the help of a (staggered) particle-hole transformation.
Since we take the non-interacting fermionic half-bandwidth $D$ as the unit of energy, the bosonic hopping amplitude $t_{b}$ and the interaction $U_{fb}$ are the remaining adjustable parameters.

\begin{figure}
\includegraphics[scale=0.28]{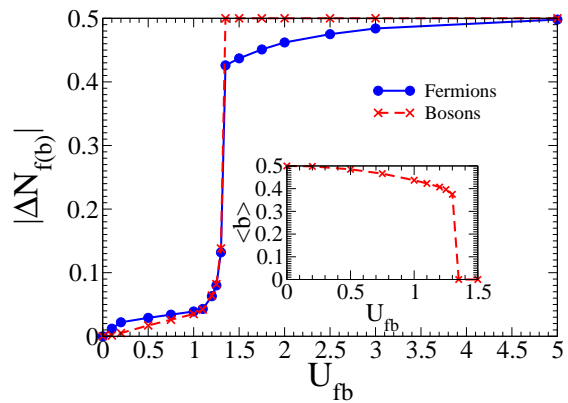}
\caption{(Color online) Amplitude of the CDW for fermions (blue circles, solid line) and hard-core bosons (red crosses, dashed line) as a function of the fermion-boson interaction 
$U_{fb}$ for the case when $zt_{b}=0.4 D$.  In the inset we plot the bosonic superfluid order
parameter as a function of the fermion-boson interaction $U_{fb}$.}
\label{CDW_vc_Ufb}
\end{figure}

Our results are shown in the $U_{fb}-t_{b}$ phase diagram in Fig. \ref{PD_hard_core}. 
For weak repulsion between fermions and bosons we obtain a {\em supersolid phase} with a small CDW amplitude(see Fig. \ref{CDW_vc_Ufb}).
For strong interactions between fermions and bosons we obtain a bosonic {\em alternating Mott insulator phase} (AMI) together with a charge density wave (CDW) of the fermions.
In this phase the fermionic CDW amplitude $|\Delta N_f|$ is almost maximal, while the bosons are completely localized and have a CDW amplitude equal to $|\Delta N_{b}|=0.5$. Taking into account virtual bosonic particle-hole excitations beyond Gutzwiller would however lead to a slightly smaller bosonic CDW amplitude. This transition is very similar to the one for bosons in a superlattice: upon increasing the potential difference between the sublattices there is a Mott-insulator transition at half filling \cite{ABH}. 
For intermediate coupling 
both solutions are stable within GDMFT. To determine which of them corresponds to the ground 
state, we have compared their energies as given by Eq.~(\ref{GSE}).  We find that the supersolid phase always has the lower energy, i.e. 
the ground state is the supersolid. The coexistence of GDMFT solutions is a strong indication for a first order phase transition (at $T=0$). As shown in  Fig. \ref{PD_hard_core}, the critical value 
$U_{fb}^{c}$ for the phase transition from the supersolid into the AMI phase 
increases with the bosonic tunneling amplitude.

\begin{figure}
\includegraphics[scale=1.0]{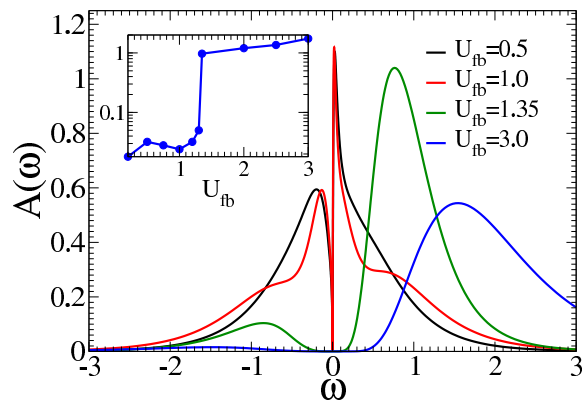}
\caption{(Color online) Fermionic spectral function in a mixture with hard-core bosons and $z t_{b}=0.4 D$ for different values of $U_{fb}$. In the inset we plot the size of the gap in units of $D$ as a function of $U_{fb}$. The gap is defined by the frequencies for which the spectral function has half its maximal value. }
\label{spectrum}
\end{figure}

The fermionic spectrum is always gapped. Spectral densities are shown in Fig. \ref{spectrum}. The gap is small for the supersolid phase, but at the transition point there is a jump in the gap and in the AMI phase it becomes of the order of the non-interacting half-bandwidth $D$ (see inset of Fig. \ref{spectrum}). This implies that the latter phase will be more stable against finite temperature effects.

So far we have considered repulsive interactions between bosons and fermions. To see what happens for attractive interactions $U_{fb}<0$ we apply a staggered particle-hole transformation to the fermions, \mbox{$c_{i} \rightarrow (-1)^{i}c_{i}^{\dagger}$}, which leads to a minus-sign in front of the Bose-Fermi interaction term. This implies that for attractive interactions we obtain the same quantum phases, but the CDW-oscillations are now in-phase, instead of out-of-phase as for repulsive interactions.


\begin{figure}
\includegraphics[scale=0.28, angle=-90]{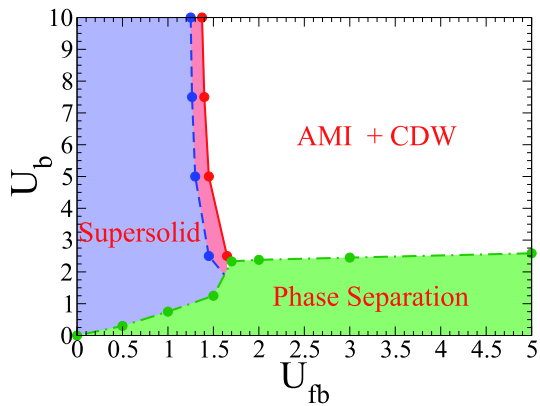}
\caption{(Color online) Phase diagram of  the Fermi-Bose Hubbard model with spinless fermions. Both fermions 
and bosons are half-filled and $z t_{b}=0.4 D$. Stable phases are the supersolid (left of the  red solid line 
and above the green dash-dotted line) and the alternating Mott insulator (AMI) phase with charge density wave (CDW) (right of the red solid line and above the green dash-dotted line).
In the area between the red solid line and blue dashed line both solutions are stable. 
Below the green dash-dotted line phase 
separation (PS) takes place. 
}
\label{PD_soft_core}
\end{figure}

We now proceed by considering finite interactions between the bosons, i.e. relaxing the hard-core condition, but still assume the fermions and the boson to be half-filled. We consider the case that the bosons are slightly slower than the fermions: $z t_{b}=0.4 D $. Our findings are summarized in the $U_{fb}-U_{b}$ phase diagram in Fig. \ref{PD_soft_core}. For strong bosonic repulsion $U_b$ the results are similar to the ones found for hard-core bosons: we find a supersolid for weak $U_{fb}$ and the alternating Mott insulator phase for stronger $U_{fb}$, separated by a first order transition.
The critical interspecies repulsion at the transition between supersolid and the AMI phase increases when the value of the bosonic repulsion $U_b$ is reduced. This is because the supersolid state acquires a lower energy when $U_b$ is decreased, whereas the energy of the AMI phase remains the same.
For weak interactions $U_b$ among the bosons, the half-filled state is unstable towards PS.  
In this parameter regime we do not find a converged GDMFT solution where the bosons and the fermions are half-filled. To establish the occurrence of phase separation we also performed calculations away from half filling. We found a pronounced jump in the density as a function of the chemical potential and coexisting solutions close to the position of the jump. 
Moreover, we observed that for strong interspecies repulsion the phase separation is always complete. 
This allowed us to compare the energies of the PS- and AMI states, which yields  
the green line as depicted in Fig. \ref{PD_soft_core}. We have checked that comparison of energies yields the same boundary for phase separation as deduced from the disappearance of a converged homogeneous GDMFT solution.

Also in this case we can infer the effect of attractive Bose-Fermi interactions by performing a staggered particle-hole transformation for the fermions. Phase separation turns then into phase separation of bosons and fermionic holes, which is equivalent to clustering of the bosonic and fermionic particles. So for weak repulsion $U_b$ among the bosons a system with attractive interspecies interaction $U_{fb}$ will maximize its density in part of the system, leaving the rest unoccupied.

In conclusion, we have studied a mixture of half-filled spinless fermions and bosons in a three-dimensional optical lattice at zero temperature. We established the presence of a supersolid at weak Bose-Fermi repulsion. For strong interspecies interaction a first order phase transition occurs towards a state where the bosons are localized and form an alternating Mott insulator. An instability towards phase separation was observed for weak interaction among the bosons. 

We thank Immanuel Bloch and Klaus Sengstock for useful discussions.
This work was supported by the German Science Foundation DFG via 
grant \mbox{HO 2407/2-1} and the Collaborative Research Center 
SFB-TRR 49.


\begin{thebibliography}{99}
\bibitem{params}
We use the NRG parameters $\Lambda = 2.0$,
$N_{\rm iter} = 60$ and $N=1000$ kept states (using $N_f$ as a good quantum number).
The bosonic number cut-off is set to $4$.
\end{thebibliography}
\end{document}